\documentclass[aps,preprint]{revtex4}

\usepackage{bm}
\usepackage{amsmath}
\usepackage{amssymb}
\usepackage{graphicx}
\usepackage{epsfig}
\usepackage{epstopdf}

\usepackage{makecell}
\begin{document}

\def\pr{\prime}
\def\be{\begin{equation}}
\def\en#1{\label{#1}\end{equation}}
\def\d{\dagger}
\def\bar#1{\overline #1}
\def\vecrF#1{\bm{|\!|}#1\bm{\rangle}}
\def\veclF#1{\bm{\langle}{#1}\bm{|\!|}}
\def\velF#1{\bm{\langle}{#1}}
\def\H{\mathcal{H}}
\def\S{\mathcal{S}}

\newcommand{\per}{\mathrm{per}}
\newcommand{\rd}{\mathrm{d}}
\newcommand{\vare}{\varepsilon }

\title{  Boson-Sampling   with non-interacting  fermions    }

\author{V. S. Shchesnovich}

\address{Centro de Ci\^encias Naturais e Humanas, Universidade Federal do
ABC, Santo Andr\'e,  SP, 09210-170 Brazil }

\begin{abstract}
 We explore  the  conditions under which  identical particles in  unitary linear networks behave as the other species, i.e.  bosons as fermions and fermions as bosons.   It is found that   the Boson-Sampling computer of Aaronson \& Arkhipov  can be implemented in an interference experiment   with non-interacting   fermions  in an appropriately entangled   state.   Moreover,  a  scheme is proposed which   simulates  the scattershot version of the Boson-Sampling computer by preparing,  on the fly, the required  entangled state of fermions    from an unentangled  one. 
\end{abstract}
\maketitle

 \section{Introduction}

 The Boson-Sampling (BS) computer of   S. Aaronson and A. Arkhipov \cite{AA}   attracts attention of physicists due to two basic pillars:  (i) simplicity, since  only passive linear optical devices, bucket detectors, and  single-photon sources are required for its experimental implementation, moreover, clearly  not sufficient for   the universal quantum computation (UQC)  \cite{KLM,BarSan,BookNC} and (ii) at the same time, due to  interference of  bosonic  path amplitudes   the output  probability distribution  of the BS computer   lies in a  higher-order  $\#$P class of computational   complexity \cite{Valiant,C,ALOP}, asymptotically inaccessible for a classical computer even to verify the result.         In more precise terms, Ref. \cite{AA}  presents  evidence that \textit{even an approximate} classical simulation of the probability distribution in the   BS computer output results in very drastic and utterly improbable consequences for the computational complexity theory.   The optical implementation of the BS  computer  is based on   completely   indistinguishable  single photons,   producing    the Hong-Ou-Mandel  dip  \cite{HOM} (see also Ref. \cite{LB}), at a unitary  network input and non-adaptive photon counting measurements at the network output.  In practical terms, with   few dozens of single  photons the BS device would be more powerful a computer than our current   computers \cite{AA}.        Several  labs have  performed the proof of principle experiments on small networks with few single  photons    \cite{E1,E2,E3,E4,E5}.  Scalability of a realistic BS computer is still under discussion \cite{ConstErr,BSscal}  due to   unavoidable  errors in any experimental realization.  Several  realistic error models  have been   analyzed \cite{R1,R2,R3,NDBS}.    Other setups    were proposed   \cite{BSIons,GBS,TBS} which could turn out to be of advantage in experimental realization.  Though  no practical problem is known to be solvable on the BS computer and   its verification poses  profound questions \cite{Gogolin,notUniform},  with the experimental efforts in this direction \cite{ExpValid,ExpComplLinOpt} and  search for  conditional  certification protocols    \cite{ZeroProb} (see also Ref. \cite{PartIndist}), such a device  would undoubtedly have a profound impact on   physics and technology.

In all discussions of  the BS computer   it is  universally  assumed that   identical bosons are indispensable.  It is usually argued that there are   profound differences between complexity of behavior  of  non-interacting   identical bosons and fermions in unitary networks:     output probabilities of completely indistinguishable bosons are given by the absolute values squared  of the matrix permanents, classically hard to compute \cite{Valiant,ALOP},  whereas output probabilities of completely indistinguishable  fermions are given  by the absolute values squared of the matrix determinants,  easily computable. Furthermore, for   fermions there are no-go theorems  \cite{ValiantMG,FLO,NonIntFerm,FermLagr,FLORev} stating that non-interacting  identical fermions with only single-mode measurements, the so-called fermionic  linear optics (FLO), do not represent any problem in terms of complexity for the classical computer. On the other hand, it was also shown that   adding  a two-mode electric charge measurement to the FLO  allows for the UQC \cite{ChargeFermUQC} (see also Ref. \cite{FLORev}).

The BS computer raises   profound questions. What is   the physical property of non-interacting indistinguishable bosons responsible for  its computational complexity? Why  non-interacting indistinguishable fermions  do not  display   a similar complexity?  Can one implement BS computer with non-identical quantum particles instead of  identical ones?  \textit{Below we show that  identical fermions  and even non-identical quantum particles   can be employed  to emulate the BS computer  distribution  via particle counting measurement at a unitary network output  when an appropriately  entangled  input state is used.     }
Moreover, we indicate a scheme emulating  with fermions the scattershot BS computer of Ref. \cite{GBS}, which requires a  non-entangled input  state     and is based on a  non-demolition  (on-off type) particle counting   at an intermediate stage.

\section{The BS computer with identical bosons}
\label{sec2}

The BS computer   proposed in Ref. \cite{AA} relies on a quantum multi-particle interference in  an unitary linear network with single photons at  input. For below, we will generalize slightly the setup and consider an arbitrary $N$-particle input with a certain number of bosons per input mode, say $n_k$ for input $k$, in a $M$-mode unitary network with the transformation matrix $U$ relating  input $a^\dag_k$ and output $b^\dag_l$ boson creation  operators
\be
a^\dag_k = \sum_{l=1}^M U_{kl}b^\dag_l.
\en{aUb}
 Let us   denote a state from a Fock basis with occupation numbers $\vec{n} = (n_1,\ldots,n_M)$  by $\vecrF{\vec{n}}$, where a subscript $a$/$b$  will indicate  the  input/output modes.  The following Fock-space  expansion of  an  input state $\vecrF{\vec{n}}_a$  in terms of the output  Fock states $\vecrF{\vec{m}}_b$   is valid \cite{AA}
 \be
\vecrF{\vec{n}}_a = \sum_{\vec{m}}\frac{\per(U[\vec{n}\,|\vec{m}\,])}{\sqrt{\mu(\vec{n})\mu(\vec{m}\,)}}\vecrF{\vec{m}}_b, \quad |\vec{m}| = |\vec{n}|=N,
\en{E1}
where $\mu(\vec{n}) = \prod_{i=1}^M n_i!$, $|\vec{n}| = n_1+\ldots+n_M$, $\per(\ldots) $ stands for the matrix permanent \cite{Minc}, and  we denote by $U[\vec{n}\,|\vec{m}\,]$ the $N\times N$-dimensional matrix obtained from  $U$ by taking  the $k$th row $n_k$ times and the $l$th column $m_l$ times (the order of rows/columns being unimportant). Expansion (\ref{E1}) considered for all inputs with $|\vec{n}|=N$ is an unitary transformation of the Fock  states corresponding to  the  unitary transformation between the respective single-particle states, an equivalent of Eq. (\ref{aUb}),
\be
|k\rangle_a = \sum_{l=1}^MU_{kl}|l\rangle_b, \quad k = 1,\ldots, M.
\en{kUl}
 By Eq. (\ref{E1}) probability  to observe an output configuration $\vec{m}$  is
\be
 p_B(\vec{m}\,|\vec{n}\,) = |{}_b\velF{\vec{m}}\vecrF{\vec{n}}_a|^2 = \frac{|\per(U[\vec{n}|\vec{m}])|^2}{{\mu(\vec{n})\mu(\vec{m}\,)}}.
\en{E2}
For single photons at  input  ($n_k\le 1$) in  the limit $M\gg N^2$  (the boson birthday paradox limit)   \cite{AA,AK}  we have $m_k\le 1$  with the  probability  $1-O(N^2/M)$. In this case     the  probability distribution  at  output of   a  network  is   given by the absolute values squared of the  matrix permanents of submatrices of $U$ which belong to the  $\#$P class of computational complexity \cite{AA}. It was   noted recently that for $M \ge N^2$    the quantum many-body correlations  of indistinguishable particles  do not vanish  in  the thermodynamic limit \cite{TLim}.

The  above account of the BS computer setup  is, however, too  idealistic: in practice, identical  photons, besides the  input   modes (spatial \cite{AA} or time-bin \cite{TBS}) that network transforms to output modes, have also  other degrees of freedom, represented by their spectral states (see   Ref.~\cite{PartIndist}).  Correspondingly,  a particle state is a tensor product of   two factors: (i) a mode state $|k\rangle$  transformed by a network and  (ii) internal state   in the  degrees of freedom not affected by the network,  say  $|\phi_\alpha\rangle$, $\alpha=1,\ldots,N$. The  mode transformation effected by such  a network must  read
\be
 a^\dag_{k,\phi_\alpha} = \sum_{l=1}^M U_{kl} b^\dag_{l,\phi_\alpha},
 \en{aUb1}
 for $\alpha =1,\ldots,N$.  For output probability to be given  by Eq. (\ref{E2}) bosons (photons) should be completely indistinguishable (for more details, consult Ref. \cite{PartIndist}). Below we explore the dependence of the probability distribution at output of an unitary linear network on the input state.


\section{ Bosons behaving as fermions and fermions as bosons }
\label{sec3}
We will deal with both  bosons and fermions and even with non-identical particles, thus  it is convenient to first define   a combined set of notations which could be  applied  in all cases. The first quantization representation for bosons and fermions  is preferable in this respect, since it  can describe also the  states of  non-identical particles.

\subsection{First quantization representation and  $\vare$-symmetrization projectors  }
\label{sec3A}

We divide particle observables   into two classes: (i) the Hilbert space $H$ of operating modes, acted on by a unitary network according to Eq. (\ref{kUl}), and (ii)   the Hilbert space $\H$ of internal degrees of freedom not affected by the network. The single-particle Hilbert space is then a tensor product $H\otimes\H$, the   single-particle states  are denoted by $|k\rangle|\phi\rangle$, $|k\rangle\in H$ and $|\phi \rangle\in \H$ (the Roman letters will be used only  for  the mode states). Vector notations will be employed for $N$-particle states, i.e.
\be
|\vec{k}\rangle \equiv \prod_{\alpha=1}^N{\!}^{\otimes}|k_\alpha\rangle,\quad |\vec{\phi}\rangle \equiv \prod_{\alpha=1}^N{\!}^{\otimes}|\phi_\alpha\rangle,\quad |\vec{k},\vec{\phi}\rangle \equiv |\vec{k}\rangle|\vec{\phi}\rangle,
\en{vec}
where $|\vec{k}\rangle\in H^{\otimes N}$ and $|\vec{\phi}\rangle\in \H^{\otimes N}$.

The states of $N$ identical particles belong to the $\vare$-symmetric subspace  of the tensor product $H^{\otimes N}\otimes \H^{\otimes N}$,  denoted by $\S_\vare \!\left\{H^{\otimes N}\otimes \H^{\otimes N}\right\}$ below.  Here   $\S_\vare $ is the projector on the $\vare$-symmetric wave-function, where for bosons $\vare(\sigma) = 1$ and for fermions  $\vare(\sigma) = \mathrm{sgn}(\sigma)$ for a permutation $\sigma$, e.g.  in $H^{\otimes N}$ it is defined as \cite{Messiah}
\be
\S_\vare  \equiv \frac{1}{N!}\sum_{\sigma}\vare(\sigma) P_\sigma, \quad  P_\sigma|\vec{k}\rangle \equiv \prod_{\alpha=1}^N{\!}^{\otimes}|k_{\sigma^{-1}(\alpha)}\rangle,
\en{S}
where the summation is over all permutations $\sigma$ of $N$ elements. The $\vare$-symmetric state vectors  will be denoted with a superscript ``$(\vare$)", e.g.,  $|\vec{k}^{(\vare)}\rangle \equiv \S_\vare |\vec{k}\rangle$ and  $|(\vec{k},\vec{\phi})^{(\vare)}\rangle \equiv \S_\vare |\vec{k},\vec{\phi}\rangle$.
Below we will use three types of the   $\vare$-symmetrization projectors: (i) acting  on  the whole space $H^{\otimes N}\otimes \H^{\otimes N}$ of $N$ particles, (ii) on  the mode space  $H^{\otimes N}$ alone, or  (iii) on the internal space $\H^{\otimes N}$ alone. These will be denoted by   $\S_\vare $, $\S_\vare \otimes I$,  and $I\otimes\S_\vare $,  respectively.    When considering only    the projector on the symmetric state we will replace ``$\vare$" by  ``$S$"  (respectfully,
by ``$A$" for the projector on the  anti-symmetric state). The same rule will apply for the $\vare$-symmetric states.

We will denote  the occupation number   (Fock) states as follows
\be
\vecrF{\vec{n}^{(\vare)}} \equiv \sqrt{\frac{N!}{\mu(\vec{n})}}|\vec{k}^{(\vare)}\rangle =\frac{1}{\sqrt{\mu(\vec{n})}}\prod_{\alpha=1}^N a^\dag_{k_\alpha}\vecrF{0},
 \en{Fock}
where  $n_1,\ldots,n_M$ are occupation numbers corresponding to the set of modes $k_1,\ldots,k_N$ (the same notations will be used for  the bosonic and fermionic creation and annihilation operators).   The state labels, e.g.  $k_1,\ldots,k_N$,  have a fixed  order in a tensor product of single-particle states to which  $\S_\vare $ is applied, i.e. in   $\S_\vare |\vec{k}\rangle$, and in a product of  creation operators, as  in Eq. (\ref{Fock}).

 We  consider $N$  identical quantum particles  in  arbitrary  internal states $|\phi_1\rangle,\ldots,|\phi_N\rangle$, thus a state $\S_\vare |\vec{k},\vec{\phi}\,\rangle$    is,  in general,   a linear superposition of the  Fock states in $\S_\vare  \{H^{\otimes N}\otimes \H^{\otimes N}\}$.    However,  the following relation, generalizing Eq. (\ref{Fock}),  is valid
\be
\frac{1}{\sqrt{N!}}\prod_{\alpha=1}^N a^\dag_{k_\alpha,\phi_\alpha}\vecrF{0} = \S_\vare |\vec{k},\vec{\phi}\rangle = |(\vec{k},\vec{\phi})^{(\vare)}\rangle.
\en{idFock}
(Eq. (\ref{idFock}) can be verified by expansion of the internal state vector $|\phi_\alpha\rangle$ in some basis, using Eq. (\ref{Fock}),  and  employing linearity of  $\S_\vare $ and  of the creation operators, i.e.   $a^\dag_{\chi} = c_1a^\dag_{\psi} +c_2a^\dag_{\phi}$ for  $|\chi\rangle = c_1|\psi\rangle +c_2|\phi\rangle$).

The following important identity between the  $\vare$-symmetrization projectors in the Hilbert space $H^{\otimes N}\otimes \H^{\otimes N}$ will be heavily used
\be
(I\otimes \S_{\vare_2})\S_{\vare_1} = (I\otimes \S_{\vare_2})(\S_{\vare_1\vare_2}\otimes I) = \S_{\vare_1\vare_2}\otimes \S_{\vare_2}.
\en{idS}
 Indeed,  noticing that $\S_\vare $ acting in  $H^{\otimes N}\otimes \H^{\otimes N}$ can be cast  as $\S_\vare  = \frac{1}{N!}\sum_{\sigma}\vare(\sigma) (P_\sigma\otimes I) (I\otimes P_\sigma)$ where $ (P_\sigma\otimes I) $ and $(I\otimes P_\sigma)$ act in $H^{\otimes N}$ and $\H^{\otimes N}$, respectively,  and  that $P_{\sigma\tau} = P_\sigma P_\tau$ and $\vare(\sigma\tau) = \vare(\sigma)\vare(\tau)$ for two permutations $\sigma$ and $\tau$ we obtain
\begin{eqnarray*}
 (I\otimes \S_{\vare_2})\S_{\vare_1}  &= &\left(\frac{1}{N!}\sum_{\sigma}\vare_2(\sigma)\left(I\otimes P_{\sigma}\right)\right) \frac{1}{N!}\sum_{\sigma^\prime}\vare_1(\sigma^\prime)\left(P_{\sigma^\prime}\otimes P_{\sigma^\prime}\right) \\
& = &\left(\frac{1}{N!}\right)^2 \sum_{\sigma}\sum_{\sigma^\prime}\vare_1(\sigma^\prime)\vare_2(\sigma^\prime)\vare_2(\sigma\sigma^\prime)\left(I\otimes P_{\sigma\sigma^\prime}\right)\left(P_{\sigma^\prime}\otimes I\right) \\
&=& \left(\frac{1}{N!}\sum_{\sigma^{\prime\prime}}\vare_2(\sigma^{\prime\prime})\left(I\otimes P_{\sigma^{\prime\prime}}\right)\right) \frac{1}{N!}\sum_{\sigma^\prime}\vare_1(\sigma^\prime)\vare_2(\sigma^\prime)\left(P_{\sigma^\prime}\otimes I\right)\\
& =&(I\otimes \S_{\vare_2})(\S_{\vare_1\vare_2}\otimes I) ,
\end{eqnarray*}
where we have used that $\vare^2(\sigma) =1$ and changed the summation   from $\sigma$ to $\sigma^{\prime\prime} = \sigma\sigma^\prime$. Note that identity (\ref{idS}) contrasts with the orthogonality property for the same space projectors, $\S_S \S_A = 0$. Identity (\ref{idS}) is  the main reason allowing   one to simulate    bosonic behavior  in a unitary linear network  using entangled   fermions and vice versa.

\subsection{Example of two quantum particles on a balanced beam splitter}
\label{sec3B}

\begin{figure}
\begin{center}
\includegraphics[width= 0.9\textwidth]{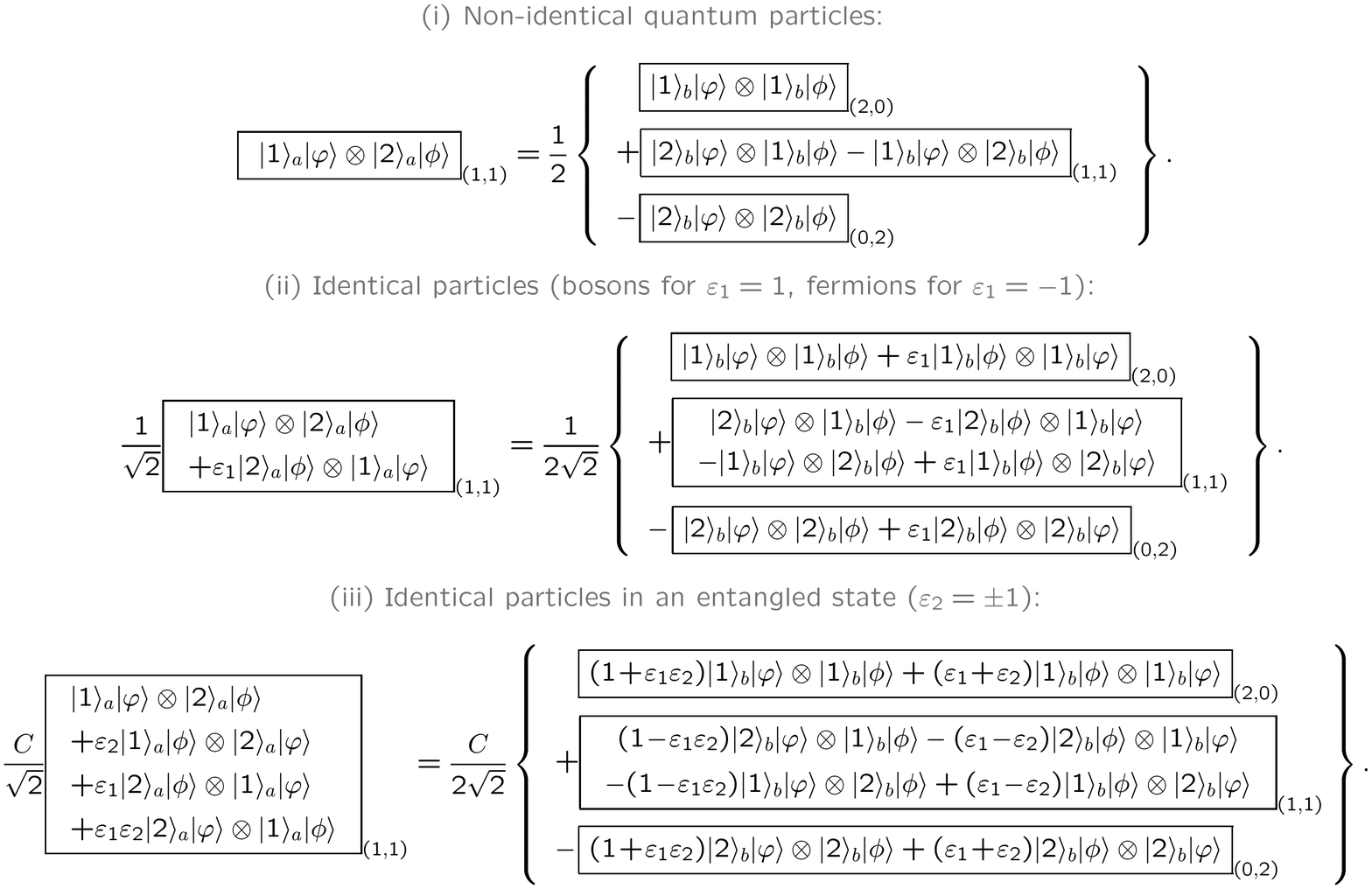}
\caption{  Diagram of output states for two particles in  internal states $|\phi\rangle$ and $|\phi\rangle$ sent through two different input  modes ($|1\rangle_a$ and $|2\rangle_a$)   of  a balanced beam splitter. Here: (i)  non-identical  quantum particles, (ii) identical bosons ($\vare_1=1$) or fermions ($\vare_1=-1$), and (iii) identical particles  in a state  symmetric  ($\vare_2=1$) or anti-symmetric  ($\vare_2=-1$)   with respect to the transposition of their internal states. Here $C = \frac{1}{\sqrt{2}}(1  + \vare_2|\langle\varphi|\phi\rangle|^2 )^{-1/2}$. Throughout, each  box gives a certain particle configuration, indicated by a subscript,  in  input ($a$, on the left) or output ($b$, on the right) modes. The input state  in  each case is  equal to  a sum of  all output  states (from all output configurations).}\label{Fig1}
\end{center}
\end{figure}
\begin{figure}
\begin{center}
\includegraphics[width=0.9\textwidth]{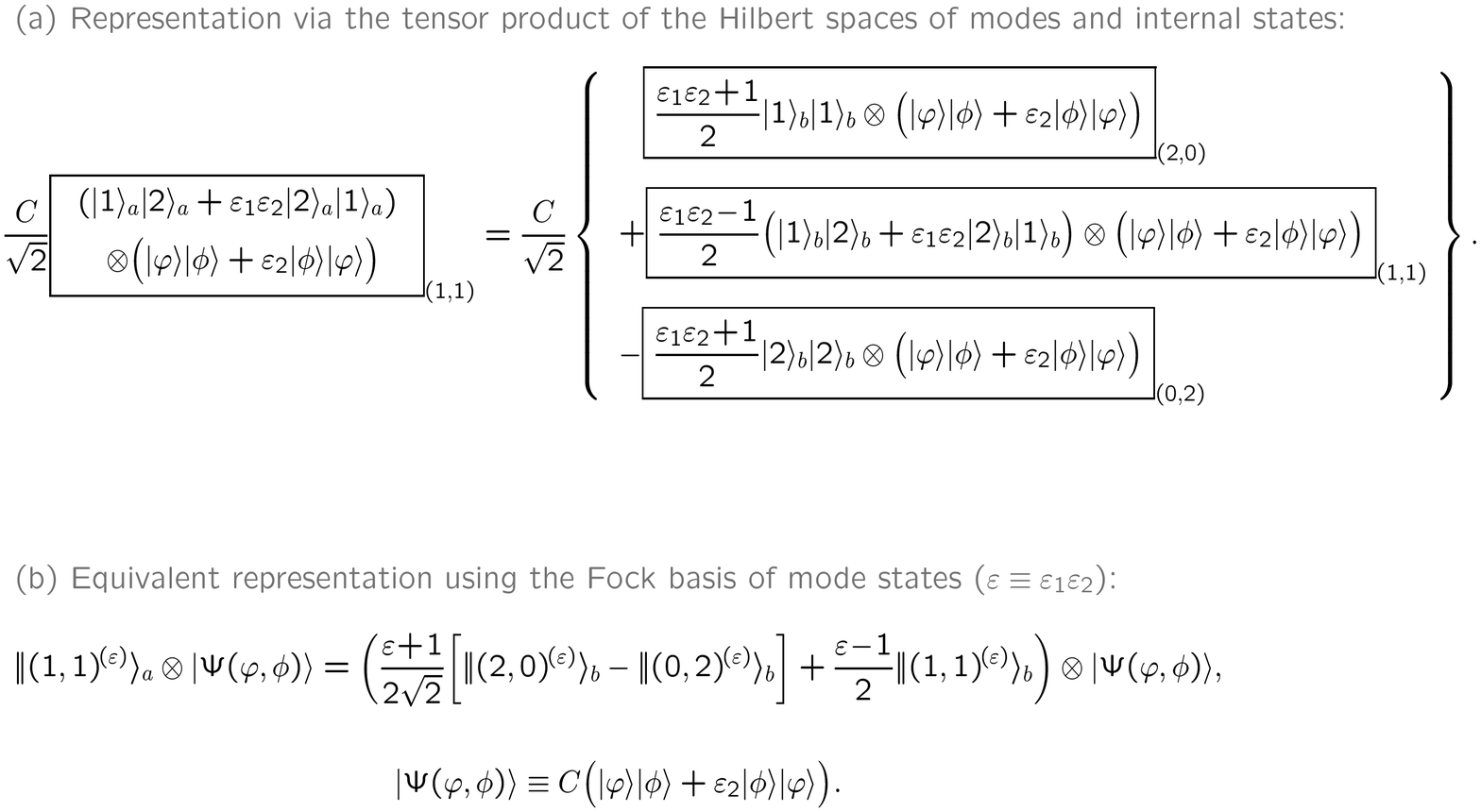}
\caption{ Case (iii) of Fig. \ref{Fig1} recast  as a tensor product of the   Hilbert spaces of modes and internal states, panel  (a), and in terms of the occupation number (Fock) states in $\mathcal{S}_{\vare}\left\{H^{\otimes N}\right\}$  with $\vare = \vare_1\vare_2$, panel (b).        }\label{Fig2}
\end{center}
\end{figure}

Before considering the general case, let us analyze in some detail the simplest possible case of two particles at a balanced beam splitter:
\be
|1\rangle_a = \frac{1}{\sqrt{2}}\left( |1\rangle_b + |2\rangle_b\right),\quad |2\rangle_a = \frac{1}{\sqrt{2}}\left( |1\rangle_b - |2\rangle_b\right).
\en{BS}
We are interested in the effect of an  input state $\vare$-symmetric in the internal degrees of freedom on the output probability distribution. The results are  presented  in Figs. \ref{Fig1} and \ref{Fig2}.   Fig. \ref{Fig1} gives the output states in the first quantization representation, where we consider (i) non-identical particles in an unentangled state, with internal states $|\varphi\rangle$ and  $|\phi\rangle$,  (ii) unentangled  identical particles in the same internal states,  and (iii) identical particles in a generally entangled  input state, $\vare$-symmetric in the internal states. In  Fig. \ref{Fig2}  we first rewrite  the state of case (iii) of Fig. \ref{Fig1}  by separating in the tensor product the state of modes    from  the internal state, panel (a), and then  cast the result in a more familiar Fock state representation   using Eq. (\ref{Fock}), panel (b). From the latter  one can easily deduce  the corresponding probabilities at the network output.  From Fig. \ref{Fig2}(b) it is seen that  the effective  bosonic behavior occurs for  $\vare  =1$ and the fermionic one for $\vare  =-1$, where $\vare = \vare_1\vare_2$, the product of  the ``natural" symmetry of the input state  with respect to the transposition of particles themselves, $\vare_1$,  and the symmetry with respect to the transposition of the internal states, $\vare_2$ (here $\vare_i$ is the value $\vare_i(\tau)$ where $\tau$ is the  transposition of two objects).
Note that  the state given in   Fig. \ref{Fig2} corresponds to the completely indistinguishable identical particles, since they behave as either indistinguishable identical bosons or fermions.

Thus, we can derive the following conclusion: an output probability of  completely indistinguishable identical particles in an unitary network  depends not   on their natural symmetry $\vare_1$ alone but on \textit{the combined symmetry} $\vare_1\vare_2$  which includes as a factor  the symmetry $\vare_2$ with  respect  to permutations of their  internal states. Below we show that this conclusion generalizes for $N$ particles  sent through an unitary network in an arbitrary input configuration.

\subsection{States $\vare $-symmetric in the internal degrees of freedom   }
\label{sec3C}

Let us consider  identical bosons first. For an output probability to be given  by Eq. (\ref{E2}) bosons  should be completely indistinguishable. One possibility is that their internal states are exactly the same $|\phi_\alpha\rangle = |\phi\rangle$, for $\alpha=1,\ldots,N$. But the latter is not the most general   state of completely indistinguishable bosons, whereas   the following  permutation-symmetric input state is \cite{PartIndist}
\be
|\Psi^{(S)}_B(\vec{n})\rangle =  \frac{c_S}{N!\sqrt{\mu(\vec{n})}} \sum_{\sigma}  \prod_{\alpha=1}^Na^\dag_{k_\alpha,\phi_{\sigma(\alpha)}}\vecrF{0},
\en{E3}
where the summation is over all permutations  $\sigma$ of $N$ elements and  $c^2_S = N!/\mathrm{per}(G)$ with $G_{\alpha\beta} \equiv \langle\phi_\alpha|\phi_\beta\rangle$.

Let us rewrite the state  of Eq. (\ref{E3})  in the first quantization representation using the  identities  of Eqs. (\ref{Fock})--(\ref{idS})
\begin{eqnarray}
\label{E5}
|\Psi^{(S)}_B(\vec{n})\rangle &=&   \frac{c_S\sqrt{N!}}{\sqrt{\mu(\vec{n})}}\left(I\otimes\S_S\right)\S_S |\vec{k},\vec{\phi}\rangle_a =  \frac{c_S\sqrt{N!}}{\sqrt{\mu(\vec{n})}}\left(\S_S\otimes\S_S\right)
|\vec{k}\rangle_a|\vec{\phi}\rangle\nonumber\\
&=& c_S \vecrF{\vec{n}^{(S)}}_a  |\vec{\phi}^{(S)}\rangle.
\end{eqnarray}
Note that $\langle\vec{\phi}^{(S)}|\vec{\phi}^{(S)}\rangle  = \langle\vec{\phi}|\S_S|\vec{\phi}\rangle = c^{-2}_S$.
From the first quantization  representation  Eq.~(\ref{E5}) is it clear  why bosons in an input  state of Eq.~(\ref{E3}) must have output  probabilities  given by Eq.~(\ref{E2}).   Indeed, the unitary  network acts in $H^{\otimes N}$, while leaving the  internal space $\H^{\otimes N}$ invariant,  therefore the output state is
obtained by using the expansion in Eq. (\ref{E1}), resulting in the output distribution given by Eq. (\ref{E2}).  The same conclusion is derived by simply noting the following mathematical identity between the symmetric states of modes, which follows from Eq. (\ref{kUl})
 \be
|\vec{k}^{(S)}\rangle_a = \sum_{ l_1\le\ldots\le l_N } \frac{\per(U[\vec{n}\,|\vec{m}\,]) }{\mu(\vec{m})}|\vec{l}^{(S)}\rangle_b.
\en{E6}
Now turning to fermions, we note that, similar to  Eq. (\ref{E6}), there is an identity relating two anti-symmetric states of modes
\be
|\vec{k}^{(A)}\rangle_a = \sum_{l_1<\ldots<l_N} \mathrm{det}(U[\vec{n}\,|\vec{m}\,]) |\vec{l}^{(A)}\rangle_b,
\en{E14}
where $k_1<\ldots <k_N$ and $l_1< \ldots < l_N$  and  the corresponding occupation numbers satisfy $n_i,m_i\le 1$~\footnote{Eq. (\ref{E14}) can be easily established by using Eq. (\ref{kUl}),  recalling the definition of the anti-symmetric state,  and using  the Laplace expansion for the matrix determinant.}. The identities (\ref{E6}) and (\ref{E14})  are equivalent forms of the  corresponding expansions for   Fock states (\ref{Fock})     of  modes,  for bosons given by Eq. (\ref{E1}),  whereas in case of  fermions  one must  replace the matrix permanent   by the matrix determinant.

Now let us  analyze the general case of an input state similar to that of Eqs.~(\ref{E3})-(\ref{E5}). Consider an   input state (bosons or fermions) of the following form
\be
|\Psi^{(\vare_2)}_{\vare_1}(\vec{n})\rangle =  c_{\vare_2}\sqrt{\frac{N!}{\mu(\vec{n})}}\left(I\otimes \S_{\vare_2}\right) \S_{\vare_1} |\vec{k},\vec{\phi}\rangle_a = c_{\vare_2}\vecrF{\vec{n}^{(\vare_1\vare_2)}}_a |\vec{\phi}^{(\vare_2)}\rangle,
\en{Psi}
where we have used  Eqs. (\ref{idS}) and~(\ref{Fock}). The normalization factor is obviously given by
\be
c^{-2}_{\vare_2}= \langle\vec{\phi}^{(\vare_2)}|\vec{\phi}^{(\vare_2)}\rangle  = \frac{1}{N!}\sum_{\sigma}\vare_2(\sigma)\prod_{\alpha=1}^N\langle\phi_\alpha|\phi_{\sigma(\alpha)}\rangle,
\en{cvare}
i.e. proportional  either to the matrix permanent for $\vare_2(\sigma) = 1$, or to the  matrix determinant  for $\vare_2(\sigma) = \mathrm{sgn}(\sigma)$.  The state of Eqs. (\ref{E3})-(\ref{E5}) is just a special case of the state (\ref{Psi}) for $\vare_1 = \vare_2 =1$. Observe the following symmetry of the state in Eq. (\ref{Psi}) due to the identity (\ref{idS})
\be
\left(P_{\sigma}\otimes I\right)|\Psi^{(\vare_2)}_{\vare_1}(\vec{n})\rangle = \vare_1(\sigma)\vare_2(\sigma)|\Psi^{(\vare_2)}_{\vare_1}(\vec{n})\rangle,\quad
\left(\S_{\vare_1\vare_2}\otimes I\right)|\Psi^{(\vare_2)}_{\vare_1}(\vec{n})\rangle= |\Psi^{(\vare_2)}_{\vare_1}(\vec{n})\rangle.
\en{symPsi}

\begin{table}[htdp]
\caption{The $\vare$-symmetric states and the  interference behavior of identical particles.}
\begin{center}
\begin{tabular}{|l|c|c|}
\hline
\diaghead{\theadfont  Diag hhhhhhhhhhhhhhhhhhhh}
{ \\ Species}{Symmetry\\ }
& $\stackrel{\hbox{Symmetric  }}{\hbox{ $\bigl[\vare_2(\sigma) = 1\bigr]$}}$  &$\stackrel{\hbox{Anti-symmetric  }}{\hbox{ $\bigl[\vare_2(\sigma) = \mathrm{sgn}(\sigma)\bigr]$}}$ \\
\hline
Bosons $\qquad \bigl[\vare_1(\sigma) = 1\bigr]$ & $\stackrel{\hbox{$|\Psi^{(S)}_S(\vec{n})\rangle = c_S \vecrF{\vec{n}^{(S)}}_a|\vec{\phi}^{(S)}\rangle $}}{\hbox{ $\bigl[$Bosonic behavior$\bigr]$}}$  &
$\stackrel{\hbox{$|\Psi^{(A)}_S(\vec{n})\rangle = c_A \vecrF{\vec{n}^{(A)}}_a|\vec{\phi}^{(A)}\rangle $}}{\hbox{ $\bigl[$Fermionic behavior$\bigr]$}}$   \\
\hline
Fermions $\bigl[\vare_1(\sigma) = \mathrm{sgn}(\sigma)\bigr]$ &$\stackrel{\hbox{$|\Psi^{(S)}_A(\vec{n})\rangle = c_S \vecrF{\vec{n}^{(A)}}_a|\vec{\phi}^{(S)}\rangle $}}{\hbox{ $\bigl[$Fermionic behavior$\bigr]$}}$   &  $\stackrel{\hbox{$|\Psi^{(A)}_A(\vec{n})\rangle = c_A \vecrF{\vec{n}^{(S)}}_a|\vec{\phi}^{(A)}\rangle $}}{\hbox{ $\bigl[$Bosonic behavior$\bigr]$}}$  \\
\hline
\end{tabular}
\end{center}
\label{default}
\end{table}

In section \ref{sec3D} we discuss  in some detail the particle counting  measurement and  show that the input state (\ref{Psi}) results in the bosonic behavior for $\vare_1\vare_2=1$ and in the  fermionic one for $\vare_1\vare_2 = \mathrm{sgn}$, the corresponding output distributions being given by the matrix permanents and by the matrix determinants, respectively.
However, the input state (\ref{Psi})  lies  not in the $\vare_1\vare_2$-symmetric  but in   the $\vare_1$-symmetric subspace $\S_{\vare_1}\{H^{\otimes N}\otimes \H^{\otimes N}\}$, i.e. it is a state of identical bosons for $\vare_1=1$ and identical fermions for $\vare_1=\mathrm{sgn}$. Indeed,  by  using  Eq. (\ref{idFock}) the state of Eq. (\ref{Psi}) can be cast also as
\be
|\Psi^{(\vare_2)}_{\vare_1}(\vec{n})\rangle = \frac{c_{\vare_2}}{N!\sqrt{\mu(\vec{n})}} \sum_{\sigma}\vare_2(\sigma)  \prod_{\alpha=1}^Na^\dag_{k_\alpha,\phi_{\sigma(\alpha)}}\vecrF{0},
\en{PsiFock}
where the creation operators are bosonic for $\vare_1 = 1$ and fermionic for $\vare_1 = \mathrm{sgn}$.   From Eqs. (\ref{Psi}) and (\ref{PsiFock}) it is seen that  for identical  particles to behave as the other species, i.e.  bosons   as fermions  and fermions as bosons, the input state  must be  anti-symmetric with respect to their internal states (the single-particle   states $|\phi_1\rangle,\ldots,|\phi_N\rangle$ must be linearly independent),  which is  an entangled state  \cite{Ghirardi,SlaterRank}.  In contrast,  the particles show their  ``natural" behavior  in  an input state  symmetric with respect to their internal states, which includes as a special case the   unentangled state of all particles being  in the same internal state. The four possible  types of input states of Eqs. (\ref{Psi}) or  (\ref{PsiFock})  of  two particle species and  two symmetries   with respect to permutations of their internal states vs. the particle behavior are given in Table~I.

Finally, we note that our method of emulating fermionic/bosonic behavior with bosons/fermions in a \textit{single network}  differs  from that of   Ref. \cite{FStatSimul} which requires   $N$ identical networks.

\subsection{Particle counting  measurement   and  output probabilities  }
\label{sec3D}

Above it was assumed that the Fock representation  $\vecrF{\vec{n}^{(\vare_1\vare_2)}}$ for  the operational modes  of an  input state  given by Eq. (\ref{Psi}) results in the same probability distribution  at a network output as that of the identical particles  with the symmetry $\vare_1\vare_2$. Since the symmetry of the Fock state does not always coincide with the ``natural'' symmetry of the identical particles this is not obvious. Let us analyze why this so in  some detail.    First,  we assume that particle detectors  are  not   distinguishing  between   various internal states (for instance, the spectral states in the case of photons).  Second, we consider  particle counting as  ideal, i.e. without false counts or particle loses.  Under these assumptions, let us derive the corresponding positive operator-valued measure (POVM) describing such  particle counting measurement, where one detects the number of particles in each output mode of a network without distinguishing between different internal states of the particles. For   $N$  identical particles,  an element $\Pi^{(\vare)}(\vec{m})$ of such a POVM  corresponding to  an output configuration $\vec{m}$ reads (see also appendix A of  Ref. \cite{NDBS} for the case of photons)
\be
\Pi^{(\vare)}(\vec{m}) = \frac{1}{\mu(\vec{m})}\sum_{\vec{j}}\left[ \prod_{\alpha=1}^Nb^\dag_{l_\alpha,j_\alpha}\right]\vecrF{0}\veclF{0}\left[ \prod_{\alpha=1}^Nb_{l_\alpha,j_\alpha}\right] = \S_\vare \Pi_{\vec{l}}\,\S_\vare,
\en{Pim}
where $|j\rangle$, $j=1,2,3,\ldots$, is a basis of the internal single-particle space $\H$, $\vec{l}$ is a vector of output modes corresponding to the  configuration
$\vec{m}$, and $\Pi_{\vec{l}}$ is defined as follows
\be
\Pi_{\vec{l}} \equiv  \frac{N!}{\mu(\vec{m})}|\vec{l}\,\rangle_b{\,}_b\langle\vec{l}\,|\otimes I.
\en{Pil}
Indeed,  using  identity (\ref{idFock}) one can prove  equivalence of  the two representations on the r.h.s. of Eq. (\ref{Pim}). The second form of the operator $\Pi^{(\vare)}(\vec{m})$ in Eq. (\ref{Pim})  follows from the fact that it is positive and non-distinguishing  between the internal states, thus   $\S_\vare|\vec{l},\vec{j}\,\rangle$ is  its eigenstate  for an eigenvalue  which depends only on the corresponding    output configuration  $\vec{m}$. The factor (eigenvalue)  $\frac{N!}{\mu(\vec{m})}$ is obtained by the following observation
\be
\sum_{|\vec{m}|=N} \Pi_{\vec{l}} = \sum_{\vec{l}} \frac{\mu(\vec{m})}{N!} \Pi_{\vec{l}} =  I\otimes I,
\en{Sum}
where the summation runs over all vectors $\vec{l}$ with   $1\le l_\alpha\le M$ varying independently  for each $\alpha$. Then from Eqs. (\ref{Pim}) and (\ref{Sum}) we obtain
\be
\sum_{|\vec{m}|=N} \Pi^{(\vare)}(\vec{m}) = \S_\vare,
\en{Sum2}
i.e. the identity operator in the $\vare$-symmetric subspace $\S_\vare \!\left\{H^{\otimes N}\otimes \H^{\otimes N}\right\}$ corresponding to  identical particles.

The output probability of a configuration $\vec{m}$ for an arbitrary input $|\Psi_\vare(\vec{n})\rangle \in \S_\vare \!\left\{H^{\otimes N}\otimes \H^{\otimes N}\right\}$  reads
\be
p_\vare(\vec{m}|\vec{n}) = \langle \Psi_\vare(\vec{n})|\Pi^{(\vare)}(\vec{m})|\Psi_\vare(\vec{n})\rangle = \langle \Psi_\vare(\vec{n})|\Pi_{\vec{l}}\,|\Psi_\vare(\vec{n})\rangle ,
\en{pmn}
where the second form  takes into account that    $|\Psi_\vare(\vec{n})\rangle$ is    a $\vare$-symmetric state (thus  the projector $\S_\vare$ in the second form of the detection operator in Eq. (\ref{Pim})  is redundant). Now, substituting the input state of Eq. (\ref{Psi}) into  Eq. (\ref{pmn})    and taking into account the symmetry (\ref{symPsi}) we obtain
\begin{eqnarray}
p_{\vare_1}(\vec{m}|\vec{n}) &= & \langle \Psi^{(\vare_2)}_{\vare_1}(\vec{n})|\Pi_{\vec{l}}\, |\Psi^{(\vare_2)}_{\vare_1}(\vec{n})\rangle
 = \langle \Psi^{(\vare_2)}_{\vare_1}(\vec{n})|(\S_{\vare_1\vare_2}\otimes I)\Pi_{\vec{l}}\,(\S_{\vare_1\vare_2}\otimes I) |\Psi^{(\vare_2)}_{\vare_1}(\vec{n})\rangle\nonumber\\
&=& |{}_b\velF{\vec{m}^{(\vare_1\vare_2)}}\vecrF{\vec{n}^{(\vare_1\vare_2)}}_a|^2,
\label{pmnPsi}\end{eqnarray}
since by  Eqs. (\ref{Fock}) and (\ref{Pil})
\be
(\S_{\vare_1\vare_2}\otimes I)\Pi_{\vec{l}}\,(\S_{\vare_1\vare_2}\otimes I) = \frac{N!}{\mu(\vec{m})}|\vec{l}{\,}^{(\vare_1\vare_2)}\rangle_b {\,}_b\langle\vec{l}{\,}^{(\vare_1\vare_2)}|\otimes I = \vecrF{\vec{m}^{(\vare_1\vare_2)}}_b{\,}_b\veclF{\vec{m}^{(\vare_1\vare_2)}}\otimes I.
\en{idPi1}

One of   conclusions to derive from  the  above analysis is this: The    probability of an output configuration $\vec{m}$ in case of identical particles in an input  state of
Eq.~(\ref{Psi})  is  the same as  of   non-identical ones in the same input state in the first quantization representation.  Indeed, if one  counts non-identical quantum particles  and then simply ``forgets" the particle labels, thus summing up   identical   output probabilities  from  all POVM elements $\Pi_{\vec{l}}$ corresponding to the same output configuration $\vec{m}$, one gets the same probability as for identical particles. We see that for emulation of the behavior of   completely indistinguishable identical particles one needs an entangled state of non-identical particles.   However, using  the state of  Eq. (\ref{Psi}) is  not the simplest way to emulate  indistinguishable  identical  particles with non-identical ones, one may as well employ  a state of the form  $\left\{ \sqrt{\frac{N!}{\mu(\vec{n})}}\S_\vare|\vec{k}\,\rangle_a   \right\} |\phi\rangle^{\otimes N}$, i.e. an equivalent of the Fock state  (\ref{Fock}).  The main problem in such an emulation  would be the process which    supplies  the required entangled state.   

\section{The scattershot BS computer with fermions}
\label{sec5}

Implementation of the BS computer with fermions requires generation of an entangled input state of  $N$ particles $|\Psi^{(A)}_{A}(\vec{n})\rangle$ of Eq. (\ref{Psi}) with the input configuration $\vec{n}$ satisfying $n_\alpha\le1$, $\alpha = 1,\ldots,N$. If  one is not able to produce such an entangled state by  means of the FLO and some particle counting measurements,   the above  boson-fermion duality would be just a curious feature with no pathway to implement the BS computer with  identical  fermions. Below we show that there is at least one  possibility to engineer  the required  entangled state of fermions   where  the key role is played by the  boson birthday paradox.

\begin{figure}[htb]
\begin{center}
\medskip
\includegraphics[width=0.85\textwidth]{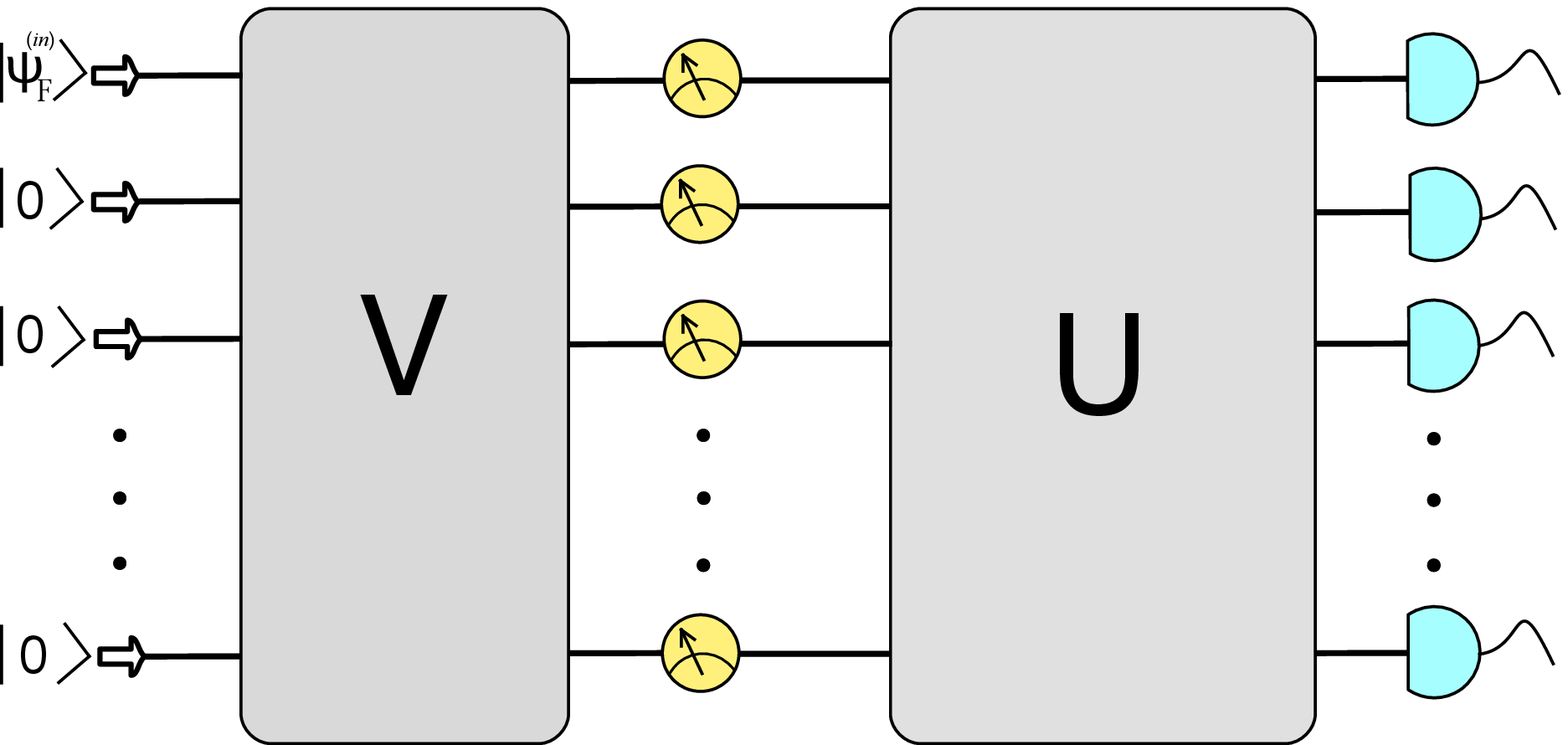}
\caption{  Schematic  depiction of     the   BS computer  with fermions. The $N$-fermion Fock state Eq. (\ref{E15}) is at  input mode $s=1$ of a $M$-mode  network $V$   with $|V_{1,k}|={1}/{\sqrt{M}}$, whereas at other input modes of $V$ is vacuum. Non-absorbing  particle    detectors, which do not  distinguish between the internal states of fermions, are placed between the network $V$ and a Haar-random network $U$.  They register which of the input modes of   $U$ contain a fermion. Particle detectors  at   output modes of the network  $U$ sample the BS output distribution averaged   over the input modes, i.e. the scheme implements the scattershot BS computer  of Ref. \cite{GBS}.     }\label{Fig3}
\end{center}
\end{figure}

 It is well-known that the FLO is  very limited in its multi-particle entangling power. For a  general $N$-particle (entangled) state there seems to be a  no-go theorem due to  a very unfavorable scaling  of  free parameters with $N$    \cite{TichyLimEnt}. But, surprisingly,  one can  prepare the needed antisymmetric entangled state   $|\Psi^{(A)}_{A}(\vec{n})\rangle$    from the following unentangled   (see Refs. \cite{Ghirardi,SlaterRank}) $N$-fermion  Fock state
 \be
|\Psi^{(in)}_{F}\rangle = \bigl[\mathrm{det}({G})\bigr]^{-\frac12}\prod_{\alpha=1}^Nd^\dag_{1,\phi_\alpha}\vecrF{0},
\en{E15}
 where the operator $d^\dag_{1,\phi_\alpha}$ creates a fermion  in some  operational mode with index $1$ and in an   internal state $|\phi_\alpha\rangle$, and ${G}_{\alpha\beta} = \langle\phi_\alpha|\phi_\beta\rangle$.    To this goal one needs to use an auxiliary $M$-mode unitary   network, whose matrix $V$ is assumed to satisfy  the Fourier type condition for the first row $|V_{1k}| = 1/\sqrt{M}$. The state of Eq.~(\ref{E15}) can be rewritten as follows $|\Psi^{(in)}_F\rangle = c_A\vecrF{\vec{q}{\,}^{(S)}}_d  | \vec{\phi}^{(A)}\rangle$, with $q_\alpha = N\delta_{\alpha,1}$  and  $c_A$ being the normalization of the internal state $| \vec{\phi}^{(A)}\rangle$ (given by  Eq. (\ref{cvare})).  According to  Eqs. (\ref{E1}) or (\ref{E6}), the state (\ref{E15}) is  transformed  by the network $V$ to a state whose  expansion over the  Fock states   $\vecrF{\vec{n}^{(A)}}_a$ (such that $|\vec{n}|=N$) reads
 \be
 |\Psi^{(in)}_F\rangle = c_A\vecrF{\vec{q}{\,}^{(S)}}_d  | \vec{\phi}^{(A)}\rangle =  c_A\sum_{\vec{n}}\sqrt{\frac{N!}{\mu(\vec{n})}}\left[\prod_{k=1}^M \left(V_{1,k}\right)^{n_k}\right]\vecrF{\vec{n}^{(A)}}_a | \vec{\phi}^{(A)}\rangle.
\en{E16}
Now assume that the network $V$ satisfies the boson birthday paradox  condition $M\gg N^2$. In this case, we have  almost surely the  Fock states on the r.h.s. of  Eq. (\ref{E16}),  at  output of the network $V$, to satisfy $n_k\le 1$.       Indeed, from Eq. (\ref{E16}) we have probability for  each particular  output with $n_k\le 1$ (i.e. $k_\alpha \ne k_\beta$ for $\alpha\ne \beta$)
\be
P(k_1,\ldots,k_N)  = N!\prod_{\alpha=1}^N|V_{1,k_\alpha}|^2 = \frac{N!}{M^N}.
\en{E17}
 We get a non-bunched state over random $N$ output modes of the network $V$  with the probability  $P_{BS}  = \sum_{\vec{k}} P(k_1,\ldots,k_N) = \frac{(M-N+1)!}{N!(M-N)!} \frac{N!}{M^N} = \prod_{q=1}^{N-1}(1-\frac{q}{M}) \approx 1 - \frac{N(N-1)}{2M}$~\footnote{We have basically repeated the boson birthday paradox derivation \cite{AA,AK}, but for a  particularly  chosen  network, not for a Haar-random  one.}.  The Fourier-type  condition    $|V_{1k}| = 1/\sqrt{M}$ corresponds to a local maximum of   $P_{BS}  $ as function of $V$.

Since the modes containing a fermion at output of the network $V$ are a random set, 
the scheme using such a network can simulate  the scattershot BS computer of Ref.~\cite{GBS}   with fermions in the input state (\ref{E15}), where the boson birthday paradox condition is required  also for  the BS computer itself \cite{AA}.  To this goal one must  detect a presence of a fermion in an output mode of the  $V$-network without  disturbing the internal $N$-particle state,  unaffected by   $V$.  This is possible, at least in principle, by using non-absorbing particle  detectors that do not distinguish between the internal states, as discussed in section \ref{sec3D}. Indeed,      non-absorbing particle detection  of fermions in output modes of $V$  is described by the POVM of Eq. (\ref{Pim}) where $\vare = \mathrm{sgn}$.      The possibility of the non-disturbing detection is guaranteed by the  following commutation rule
\be
\left(I \otimes \S_{A}\right)\Pi^{(A)}(\vec{m}) = \Pi^{(A)}(\vec{m})\left(I \otimes \S_{A}\right),
\en{E20}
which can be easily established  from the second form of $\Pi^{(A)}(\vec{m})$  on the r.h.s. of Eq. (\ref{Pim})~\footnote{Here we note that an identity  similar to Eq. (\ref{E20}) is valid for bosons. More generally, in both cases,  $\Pi^{(\vare_1)}(\vec{m})$ commutes also with   $I \otimes \S_{\vare_2}$ due to    the second form of $\Pi^{(\vare_1)}(\vec{m})$  on the r.h.s. of Eq. (\ref{Pim}).}.

Therefore, we have at least one  scheme for  implementation of the BS computer with identical fermions, depicted  in  Fig. \ref{Fig3}. In one respect the scheme of Fig. \ref{Fig3} is very similar to the   scattershot  BS computer with photons proposed  in Ref.~\cite{GBS}, where an  additional averaging over random, but  known in each run,  non-vacuum input modes of the network is implemented.  Indeed,   the BS computer with fermions  of   Fig. \ref{Fig3} has almost surely   $N$ non-vacuum  input modes at  network $U$, each containing a single fermion. The indices of the non-vacuum input modes  constitute a random set   \textit{known in each run}, where each particular configuration is  generated with the same  probability. The fact that the  non-vacuum  mode indices of the network $U$ are known in each run leads to the same complexity of its  output  probability distribution  as  of  the original  BS computer with a fixed set of non-vacuum input modes \cite{AA,GBS}.

\section{Conclusion}
\label{sec6}
We  have found the   boson-fermion  duality   in unitary linear networks, i.e. bosons behaving as fermions and fermions as bosons in the appropriately entangled  input states,  allowing for  the BS computer  with non-interacting fermions.  Such a possibility  provides an  insight on  the physical origin  of its computational complexity. Indeed, it was previously believed that only  non-interacting identical bosons posses a fundamental   feature  allowing one to implement the BS computer.  It is  shown here that one can substitute  bosons  with non-interacting identical  fermions   in  the   antisymmetric entangled state over  their internal degrees of freedom not affected by a network. Moreover, such an entangled $N$-particle  state can be engineered   by means of  the FLO  and a non-absorbing  multi-mode particle counting measurement from  a state of  $N$ fermions sharing a common (operational) mode and   distributed  over linearly-independent   internal degrees of freedom.   This agrees with the  previous result \cite{ChargeFermUQC}  that the FLO can be promoted to a higher computational complexity, in our case to   sampling from  a \#P-class  problem, by using   multi-mode non-absorbing particle counting.

Our goal   was a proof of principle of the BS computer with fermions,   an experimental implementation    using currently available technology  is  challenging. However, there is an important progress in this direction --  recent observation of the Hong-Ou-Mandel effect with massive identical particles, with fermions (electrons) \cite{add1} and with bosons (Helium atoms) \cite{add2}.   We have  not touched upon in  our  discussion      which degrees of freedom of fermions   could serve as their  internal modes and, respectively, which  would be the operational modes   transformed by a  network (and how to build such a network).  Such  questions and practical ways to implement the  BS computer with fermions are left for the  future research (in this respect, there are now already  two different implementations for the BS computer with photons \cite{AA,GBS} and   one with excitation quasiparticles in trapped ions \cite{BSIons}).  As a byproduct, we have also found how to implement  the recently experimentally demonstrated method of emulation of the Fermi-Dirac statistics with bosons \cite{FStatSimul}    in a single network, instead of $N$ identical ones for $N$ particles,    by  entangling the internal states of particles instead of the operating modes transformed by a network.

Finally, from our discussion it  follows that one can also simulate the behavior of  identical particles with non-identical ones.  For instance, the famous Hong-Ou-Mandel effect can be observed with non-identical particles in a single network, if the input state is a properly entangled one.

\section{Acknowledgements}
The author is indebted to an unknown Referee for many helpful suggestions. This work was supported by the CNPq of Brazil.

\end{document}